\documentstyle[aps,multicol,
epsfig]{revtex}

\begin{document}

\title{Turbulence and passive scalar transport in a free-slip surface}
\author{Bruno Eckhardt and J\"org Schumacher 
\thanks{Present address: P.O. Box 208284,
Yale University, New Haven CT 06520-8284}}
\address{Fachbereich Physik, Philipps--Universit\"at Marburg,
         D-35032 Marburg, Germany}  
\date{\today}  
\maketitle 

\begin{abstract}     

We consider the two-dimensional (2D) flow in a flat free-slip surface that
bounds a three-dimensional (3D) volume in which the flow is turbulent.  The
equations of motion for the two-dimensional flow in the surface are neither
compressible nor incompressible but strongly influenced by the 3D flow
underneath the surface.  The velocity correlation
functions in the 2D surface and in the 3D volume scale with the same exponents.
In the viscous subrange the amplitudes are the same, but in the inertial
subrange the 2D one is reduced to $2/3$ of the 3D amplitude.  The surface flow
is more strongly intermittent than the 3D volume flow.
Geometric scaling theory is used to derive a relation between the scaling of the
velocity field and the density fluctuations of a passive scalar advected on the
surface.
\end{abstract}    

\begin{multicols}{2}
\section{Introduction}
We consider flows in a flat two-dimensional (2D) surface that bounds a
three-dimensional (3D) volume with turbulent fluid motion.  The boundary
conditions are that of a free-slip surface so that the normal velocity component
vanishes but the parallel components are not further constrained.  To some
extent, this is the situation of surface currents on a river or the sea, if
waves and ripples are absent or can be neglected.  Particles floating on the
surface reflect the properties of the flow and provide an easy visualization.
These flows have an obvious connection to oceanographical
applications\cite{Stommel,Davis}, but they apparently have not been studied in further
detail.  Even in the 
recent theoretical and experimental investigations of
the statistical properties of the particle distribution by Ott and co-workers
\cite{Ott1,Ott2,Ott3,Sommerer} the modelling was based on random dissipative
maps and not on the underlying flow.  
Similarly, Saichev and co-workers \cite{Saich97,Saich98} based their
investigation of passive particle advection and cluster formation
on Gaussian random velocity fields, white in time.
Thus, one of our aims here is to analyze the
properties of surface flows arising from Navier--Stokes dynamics
and to connect them to the statistics of
particles floating on the surface, along the lines of our previous work on
passive scalars advected in two-dimensional turbulent flows \cite{ES99}.

The flow in the surface is 
two-dimensional, but it has properties that
are different from that of the usual two-dimensional incompressible
Navier-Stokes turbulence.  Obviously, the velocity field is not constrained by
mass conservation in the surface:  there can be up- and down-welling motions in
the (incompressible) bulk which on the surface will appear as sources and sinks
for the velocity field.  Velocity and vorticity can be exchanged with the bulk
flow underneath, so that in the inviscid limit without forcing neither kinetic energy
nor enstrophy are conserved.  Such effects of compressibility arise also in
experiments in two-dimensional turbulence in soap films and were discussed
recently \cite{Mar98,Riv98,Rut98}.

The experiments of Goldburg {\em et al.}  \cite{Gol00} are close to a laboratory
realization of the kinds of flows that are investigated here.  A vertically
oscillating grid in a tank of water is used to produce turbulence.  If the water
surface is sufficiently far away from the grid it remains essentially flat and
the surface flow can be visualized with mushrooom spores.  The measured
statistical properties of the flow are close to the ones that we will derive
here.  This opens the way to further experimental studies of the statistical properties of
the velocity field and of the particle dynamics in this interesting flow.

Finally, we should like to point out that the flows are also of interest from a
theoretical point of view, since they can be thought of as flows with a symmetry
plane:  let the surface be $z=0$ and consider the reflection symmetry that under
$z$ goes to $-z$ the $z$-component of the velocity field changes sign.  This is a
symmetry of the Navier-Stokes equation, 
that is to say, if initial conditions and
driving preserve this symmetry so does the time evolved flow.

It is our aim here to derive the equations of motion for such a flow (section
IIA), to discuss the correlation function if the 3D flow is turbulent (section
IIB), to present numerical 
results on the statistics of the velocity, vorticity, and
divergence fields and on the boundary layer thickness (section III) and
to derive a relation between the fractal dimension and the velocity correlation
function for the advection of scalars within geometric scaling theory (section
IV).  Concluding remarks are given in section V.

\section{The two--dimensional flow in a free-slip surface}

In order to arrive at the properties of such a flow, two approaches are
possible:  one relies on an explicit representation of the flow field with
proper boundary conditions and the other seeks to 
derive the equations of motion from the 3-d Navier-Stokes equation.
They provide complementary information on the system.

\subsection{Flow with reflexional symmetry}

We begin with the equations of motion and the effects of symmetry.
Let $u$, $v$, $w$, be the $x$-, $y$-, and $z$-components of the velocity
field (Fig.~1) and let $p$ be the pressure field.
The surface flow can be realized as flow in a symmetry plane,
e.g. the plance $z=0$ if the velocity field is invariant under
the symmetry $(u,v,w)\rightarrow (u,v,-w)$ when
$z\rightarrow -z$. 
\begin{figure}
\begin{center}
\epsfig{file=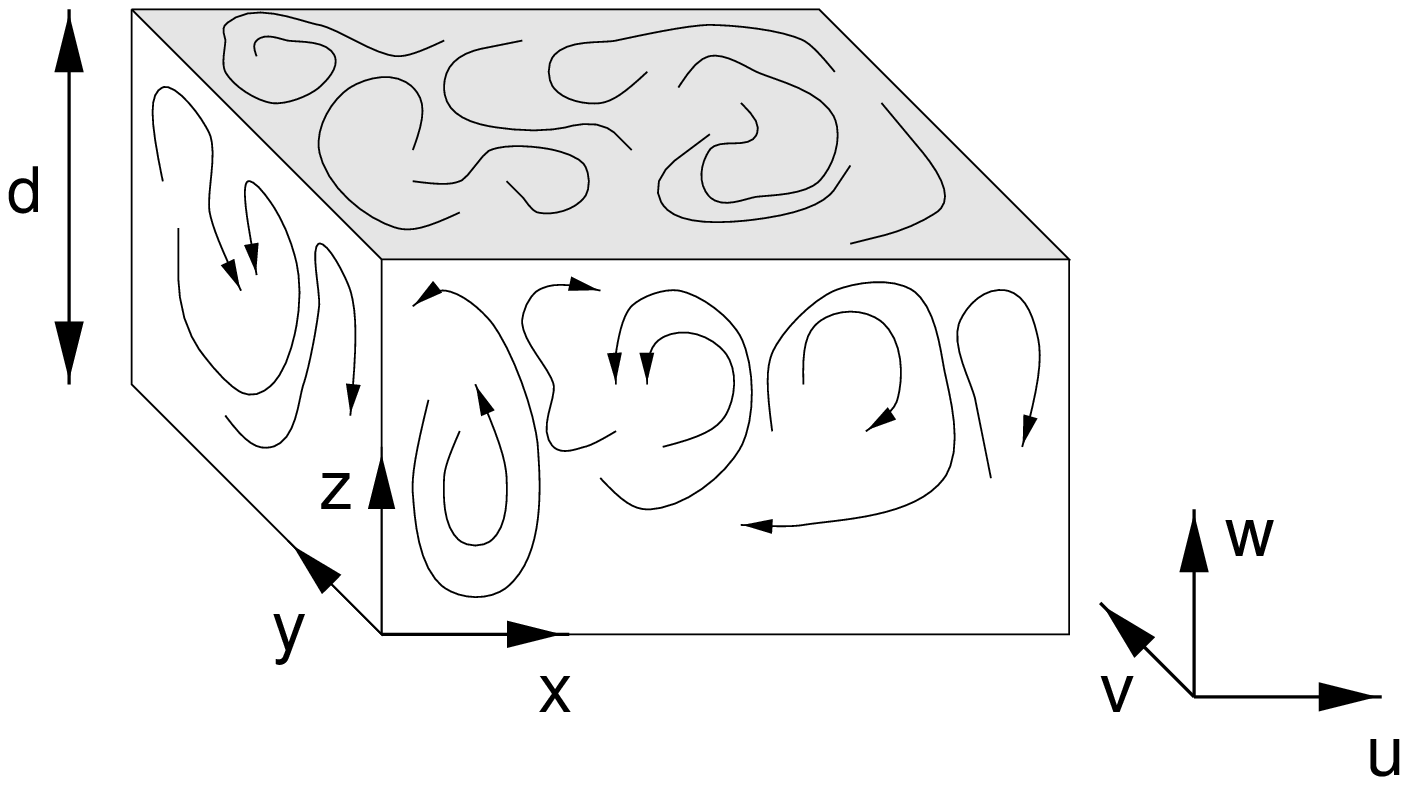,width=7.5cm}
\end{center}
\end{figure}
FIG. 1. Flow geometry. The surface flow lives in the shaded surface
above a turbulent bulk flow. $x$ and $y$ are 
the coordinates in 
the surface and $z$ is the one normal to it. In the numerical
simulation the flow is driven by a shear flow
in the $x$-direction with variations in $z$.

\vspace{0.5cm}
\noindent
This suggests to expand the velocity components in power series in $z$, with only odd powers
for $w$, and only even ones for $u$, $v$ and $p$,
\begin{eqnarray}
u(x,y,z,t) &=& \sum_{n=0}^\infty u_{2n}(x,y,t) z^{2n}\,,
\label{series1}\\
v(x,y,z,t) &=& \sum_{n=0}^\infty v_{2n}(x,y,t) z^{2n}\,,\\
w(x,y,z,t) &=& \sum_{n=0}^\infty w_{2n+1}(x,y,t) z^{2n+1}\,,
\label{series3}
\end{eqnarray}
and
\begin{equation}
p(x,y,z,t) = \sum_{n=0}^\infty p_{2n}(x,y,t) z^{2n}\,.
\end{equation}
Substitution into the Navier-Stokes equation and ordering with 
respect to powers of $z$ gives for the two main components of 
interest, $u_0(x,y,t)$ and $v_0(x,y,t)$, the equations
\begin{eqnarray}
\partial_t u_0 + (\hat{\bf u}_0\cdot\hat{\nabla}) u_0
&=& - \partial_x p_0 + \nu \hat{\Delta} u_0 + 2\nu u_2 + f_u\,,
\label{eq_u0}\\
\partial_t v_0 + (\hat{\bf u}_0\cdot\hat{\nabla}) v_0
&=& - \partial_y p_0 + \nu \hat{\Delta} v_0 + 2 \nu v_2 + f_v\,.
\label{eq_v0}
\end{eqnarray}
The hat superscripts on position vectors ${\bf x}$ and ${\bf R}$,
on the velocity field ${\bf u}$, on the gradient $\nabla$ and on the 
Laplace operator $\Delta$ indicate that they are restricted to the 
components $x$ and $y$ that lie in the surface. 
The driving of the turbulence is modelled by a volume
force with components $f_u$, $f_v$ and $f_w$; as usual we 
expect that the statistical properties of the flow depend
only weakly (through intermittency) on the kind of driving as long as it
is confined to large scales. $\nu$ is the kinematic viscosity of the fluid.

For the full 3D velocity field mass conservation ${\bf\nabla\cdot u}=0$ connects
the normal and tangential components, viz.
$\partial_z w = - \partial_x u - \partial_y v$\,
or, on the levels of the components [cf. (\ref{series1})--(\ref{series3})],
\begin{equation}
w_{2n+1} = -(\partial_x u_{2n} + \partial_y v_{2n})/(2n+1)\,.
\label{Conti}
\end{equation}
Similarly, the pressure has to be determined from the 3D relation 
$\Delta p = - \nabla\cdot[({\bf u}\cdot\nabla) {\bf u}]$.
With the power series expansion from above this becomes to 
leading order in $z$,
\end{multicols}
\begin{equation}
\hat{\Delta} p_0(x,y)+2 p_2(x,y)=
2\left[(\partial_x u_0)(\partial_y v_0)-(\partial_y u_0)(\partial_x v_0)\right]
-2 \left( \partial_x u_0 + \partial_y v_0\right)^2\,.
\label{pressure}
\end{equation}
\begin{multicols}{2}
Equations (\ref{eq_u0}), (\ref{eq_v0}) and (\ref{pressure}) 
are the equations of motion for the surface flow.
Note that besides the surface velocity field $(u_0, v_0)$ and
the surface pressure $p_0$ there are additional contributions
from higher order terms in the power series in $z$:
the viscous driving terms 
$\nu u_2$ and $\nu v_2$ from shear effects in the normal
direction and a contribution $(\hat{\Delta})^{-1} p_2 $ to the
pressure, also resulting from pressure variations in the
wall normal direction. From the point of view of the
flow in the surface, these terms are externally given and
can hence be included in the volume
driving forces. Note, however, that now the driving is no longer
confined to large scales, as assumed in the usual scaling analysis.
With all unspecified terms absorbed into effective volume forces
$\tilde f_u$ and $\tilde f_v$, the equations
of motion for $u_0$ and $v_0$ become finally
\begin{eqnarray}
\partial_t u_0 + (\hat{\bf u}_0\cdot\hat{\nabla}) u_0
&=& - \partial_x p_0 + \nu \hat{\Delta} u_0 + \tilde f_u\,,\label{eq5}\\
\partial_t v_0 + (\hat{\bf u}_0\cdot\hat{\nabla}) v_0
&=& - \partial_y p_0 + \nu \hat{\Delta} v_0  + \tilde f_v\,,\label{eq6}
\end{eqnarray}
The equations are completed by (\ref{pressure}) with $p_2=0$
for the pressure.

These equations have unusual properties. For instance, 
dotting with $\hat{\bf u}$ and integrating over a 2D volume,
the energy is not conserved in the Eulerian
limit where viscosity and driving are absent. With the local energy density,
\begin{equation}
E(x,y,t)=(u_0^2 + v_0^2)/2\,,
\end{equation}
and using eq.~(\ref{Conti}) for $n=0$ the global energy balance reads
\begin{equation}
\partial_t \langle E(x,y,t)\rangle_S = 
-\langle w_1(x,y,t) (E(x,y,t)+p_0(x,y,t)) \rangle_S,
\label{e-ave}
\end{equation}
where $\langle\cdot\rangle_S$ denotes the average over the surface $S$. 
Thus energy is permanently put in and taken out according to the 
gradients of the $z$-component of ${\bf u}$ and the pressure fluctuations.
Over large time intervals one can expect that a flow equilibrium 
with constant average energy is established and that the 
time average of the right hand side of (\ref{e-ave}) vanishes. 
It seems that the lack of energy conservation on short times gives rise
to larger fluctuations and larger intermittency corrections
(see below and \cite{Gol00}). A similar discussion applies to
the vorticity, and will be given in section IIIB below.

\subsection{Direct representation of a stress-free surface}

The alternative approach mentioned above starts from an
explicit representation of the 3D velocity field that takes the 
boundary conditions into account. Consider the Fourier 
expansion of the velocity field,
\begin{eqnarray}
u(x,y,z,t) = \sum_{\hat{\bf K},n} u_{\hat{\bf K},n}(t)
\exp(i\hat{\bf K}\cdot\hat{\bf x}) \cos(n\pi z)\,,
\label{fou1}\\
v(x,y,z,t) = \sum_{\hat{\bf K},n} v_{\hat{\bf K},n}(t)
\exp(i\hat{\bf K}\cdot\hat{\bf x}) \cos(n\pi z)\,,\\
w(x,y,z,t) = \sum_{\hat{\bf K},n} w_{\hat{\bf K},n}(t)
\exp(i\hat{\bf K}\cdot\hat{\bf x}) \sin(n\pi z)\,,
\label{fou2}
\end{eqnarray}
where the summation
extends over all 2D wave vectors $\hat{\bf K}=(K_x, K_y)$ in the surface and all
integers $n$. The sine and cosine terms take into account the 
stress-free boundary conditions at the top and bottom surface,
\begin{equation}
\partial_z u=\partial_z v=w=0 
\quad\text{for}\quad z=0 \quad\text{and}\quad z=1\,,
\label{bound}
\end{equation}
Incompressibility requires
\begin{equation}
i K_x u_{\hat{\bf K},n} + i K_y v_{\hat{\bf K},n} + n \pi w_{\hat{\bf K},n} = 0\,.
\end{equation}
One advantage of this representation is that it quickly leads
to a prediction for the two-point correlation functions. 
In the 3D case Kolmogorov scaling without intermittency gives
for the inertial regime a decay of amplitudes
$|{\bf u}_{\hat{\bf K},n}|^2 \propto |\hat{K}^2+(n\pi)^2|^{-11/3}$
\cite{Frisch}. 
In the surface, the 2D amplitudes are obtained by summation on $n$.
This brings in a factor of $K$ that compensates the one missing
from the volume element, which is $K\,dK$ in 2D rather than $k^2\, dk$
as in 3D. As a net result scaling of the correlation function does not
change. 
However, the absence of 
the third component of the velocity field reduces the 
amplitude to two third of its three-dimensional value.
For the second order structure function, defined as
\begin{eqnarray}
S_{2}(R)=
\langle|{\bf u}({\bf x}+{\bf R})-{\bf u}({\bf x})|^2\rangle\,,
\end{eqnarray}
we expect in the inertial regime
\begin{equation}
\hat{S}_2({R}) = \frac{2}{3} S_2(R) \sim R^{2/3} \,,
\label{form}
\end{equation}
where again the hat distinguishes the 2D surface from the 
3D bulk structure function.

\section{Numerical simulations}

The numerical simulations are based on a 
nearly homogeneous turbulent shear flow
bounded by stress-free surfaces at $z=0$ and $z=1$ as given in
Eq.~(\ref{bound}).  The velocity field is decomposed as in
Eqns.~(\ref{fou1})--(\ref{fou2}) and the Navier-Stokes equations are integrated
using a pseudospectral method \cite{See96,Schu00}.  The simulations were done
for Taylor Reynolds numbers $Re_{\lambda}=59$, 79 and 99, calculated from the
streamwise velocity component $u$, i.e.
$Re_{\lambda}=u^2_{rms}/[\langle(\partial_x u)^2\rangle^{1/2}\nu]$ with root  
mean square velocity $u_{rms}=\langle u^2\rangle^{1/2}$.
The
properties of the 3D bulk flow are included here only to the extent that they
are needed for the comparison between bulk and surface; they are further
analyzed in \cite{Schu00}.

Kolmogorov length $\eta=(\nu^3/\epsilon)^{1/4}$,
velocity $v_{\eta}=(\epsilon\nu)^{1/4}$, and time scales
$\tau_{\eta}=(\nu/\epsilon)^{1/2}$ 
are calculated from
the 3D energy dissipation rate in the surface, i.e.
\begin{equation}
\epsilon=\nu\sum_{i,j=1}^2\langle(\partial_i u_j)^2\rangle_S+
            \nu\langle (\partial_3 u_3)^2\rangle_S\,,
\label{epsilonsurface}            
\end{equation}
where indices 1, 2, and 3 correspond to $x$, $y$, and $z$,
respectively. For $Re_{\lambda}=99$
this dissipation rate in the surface is about $40\%$ of the 
value in the bulk.

\subsection{Structure functions of the velocity field}

Form factors in the middle of the cell and on the surface are determined from
114 statistically independent snapshots of the turbulent flow.
We focus on the scaling of the $n$-th order longitudinal structure functions, 
defined as
\begin{eqnarray}
\hat{S}_{n}^L(\hat{R},z_0)=
\langle|[\hat{\bf u}(\hat{\bf x}+\hat{\bf R},z_0)-\hat{\bf u}(\hat{\bf x},z_0)]
\cdot\hat{\bf R}/\hat{R}|^n\rangle\,.
\end{eqnarray}
In the bulk and without intermittency corrections the second
order structure function is expected to scale like $R^2$ in the viscous 
subrange and like $R^{2/3}$ in the inertial subrange \cite{Frisch}. 
A comparison between bulk and surface structure functions is shown
in Fig.~2 for $Re_{\lambda}=99$. 
\begin{figure}
\begin{center}
\epsfig{file=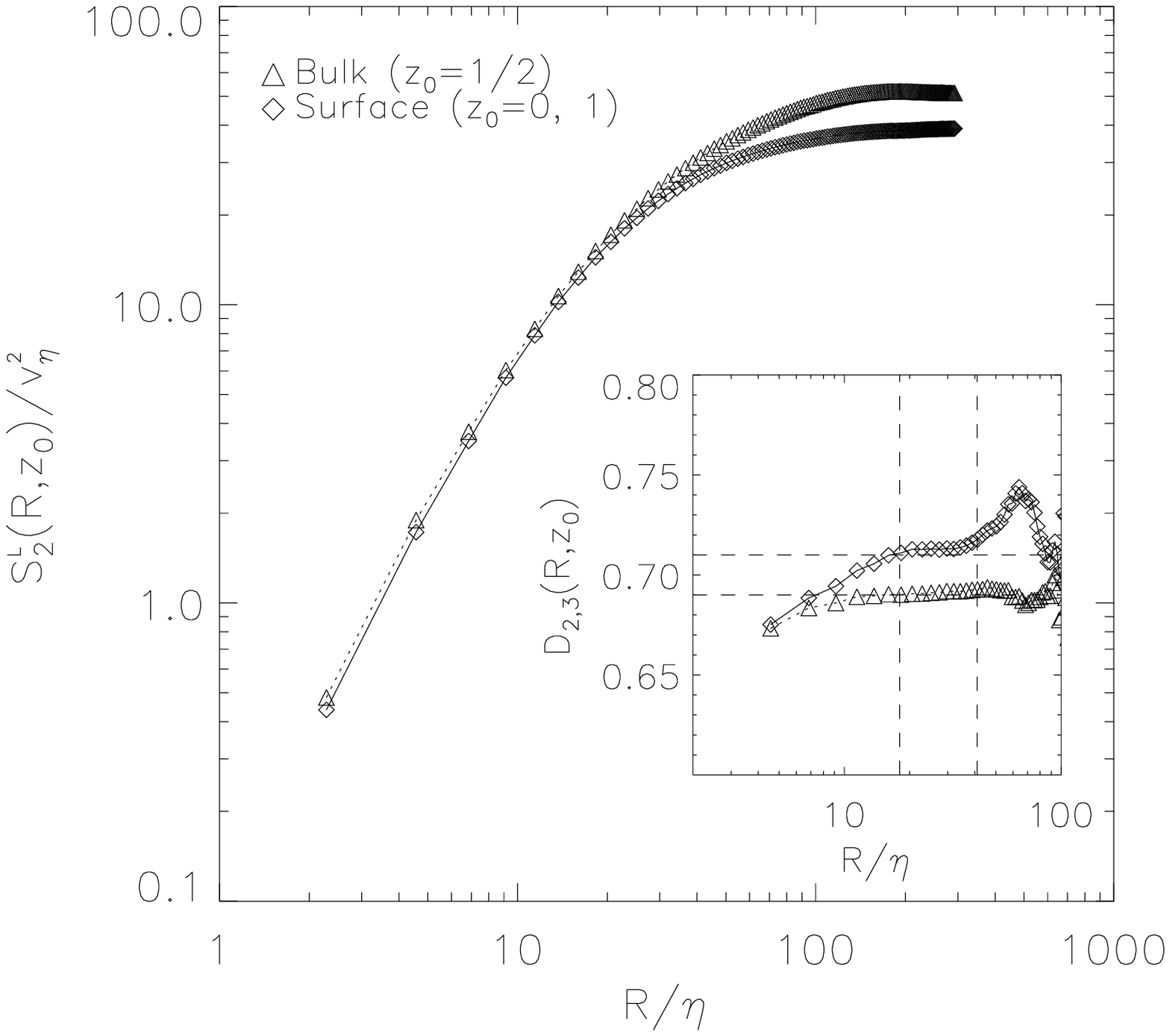,width=7.5cm}
\end{center}
\end{figure}
FIG. 2. Second order structure functions $\hat{S}_2^L(\hat{R},z_0)/v^2_{\eta}$
for $Re_{\lambda}=99$ normalized with $v_{\eta}=(\epsilon \nu)^{1/4}$ and
$\epsilon$ from (\ref{epsilonsurface}).
Data from the surfaces at $z_0=0$ and $z_0=1$ 
are indicated by diamonds and connected by continuous lines. Data in
the bulk were taken at $z_0=1/2$ and are indicated by triangles and
dashed lines. The inset shows the local scaling exponents from an 
extended self-similarity (ESS) analysis. Estimates between the 
vertical dashed lines, where the exponents are reasonably constant,
give  mean scaling exponents of 0.69 in the bulk and of 0.71
in the surface. These values are indicated by horizontal
dashed lines.
\vspace{0.5cm}
\noindent
The two structure functions
coincide in the viscous subrange but differ in the inertial subrange.
This difference is predominantly in the amplitude and not in the 
scaling exponents, and consistent with (\ref{form}). A local scaling
exponent can be defined as 
$\zeta(\hat{R})=\mbox{d}\log \hat{S}_{n}^L(\hat{R},z_0)/
\mbox{d}\log \hat{R}$.
Unfortunately, both structure functions 
do not show an algebraic scaling behavior at 
intermediate scales between the viscous and the forcing scale range for
the values of $Re_{\lambda}$ achieved here.
Therefore, we apply the extended self-similarity (ESS)
analysis \cite{benzi} to the data. A local 
ESS scaling exponent can be calculated by relating local scaling exponents
of second and third order structure functions, 
\begin{eqnarray}
D_{2,\,3}(\hat{R},z_0)=\frac{\mbox{d}\log[\hat{S}_{2}^L(\hat{R},z_0)]}
                            {\mbox{d}\log[\hat{S}_{3}^L(\hat{R},z_0)]}\,.
\end{eqnarray}
The distance vector $\hat{\bf R}$ is taken in planes of fixed $z_0$.
As shown in the inset in Fig.~2 the bulk
data give a local scaling exponent of about $0.69$, in agreement with other
observations, but  in the surface the local slope is larger, about $0.71$. 
This difference is small but statistically significant. 
Local exponents, based on averages over planes parallel to the 
surface, show almost no variation in the center of the cell but a
clear 
trend when approaching the surface.
This is demonstrated in Fig.~3 for the deviations 
$\delta D_{n,\,3}(R,z_0)=D_{n,\,3}(R,z_0)-n/3$ from classical K41-scaling for 
orders $n=2$ to $n=6$ for different $z_0$. The plane $z_0=1/2$ defines the
middle between both free surfaces. 
Data sets for two different 
Taylor-Reynolds numbers 
$Re_{\lambda}=99$ (a) and $79$ (b) are shown.
The transition from bulk to surface behavior 
can be used to define a surface layer, as discussed further in section~\ref{layer}.  
\begin{figure}
\begin{center}
\epsfig{file=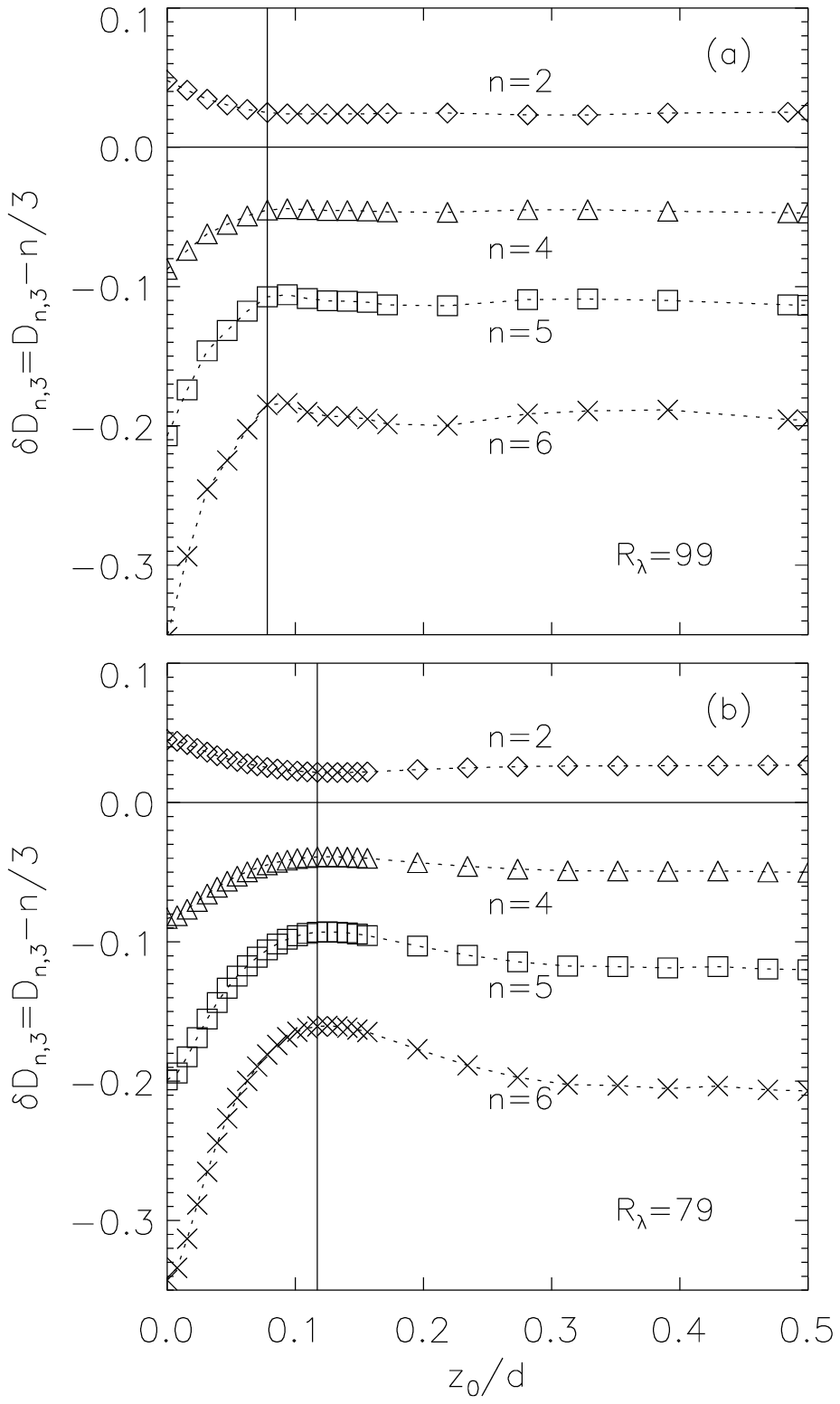,width=6.5cm}
\end{center}
\end{figure}
FIG. 3. Deviations of the local ESS scaling exponent from the classical Kolmogorov 
scaling for different heights $z_0$ of the averaging plane
and for different orders $n=2$ to $n=6$. Part (a) for
$Re_{\lambda}=99$, averaged over 228 samples. The deviations
$D_{n,\,3}(R,z_0)$ are the mean obtained 
for scales $R$ between $18\eta$ and $41\eta$, as indicated 
by vertical lines in the inset of Fig~2.
Part (b) for $Re_{\lambda}=79$, averaged over 254 samples. Here 
the exponents are obtained 
for scales $R$ between
$10\eta$ and $27\eta$. 
The vertical lines indicate the surface layer that is analyzed further in
section \ref{layer}.
\vspace{0.5cm}
\noindent

The increase in intermittency seems to be connected with an increase in
fluctuations due to lack of incompressibility and lack of energy conservation.
It is in line with results for passive scalar transport in models with
compressible Gaussian random flows that are $\delta$-correlated in time
\cite{Gaw00,Celani99} and with direct numerical simulations
\cite{Porter98,Pety00} of isotropic supersonic turbulence.  Interestingly, in
the latter case the authors also noted a strong difference to incompressible
turbulence near the crossover to the viscous subrange.  In their case vortex
filaments of high intensity and narrow regions of strong negative divergence,
due to small-scale supersonic shocks, appeared.  In our situation it
is the fluctuations due to normal shear and normal velocity components below the
surface that have a strong effect near the crossover to the viscous subrange.

In the viscous subrange the amplitudes of the structure functions agree, but in
the inertial subrange the surface structure function is smaller by a factor of
$2/3$.  In the previous section we explained the reduction in amplitude in the
inertial range by the reduction in the number of active degrees of freedom or
Fourier modes.  In the viscous subrange this argument does not apply, since we
absorbed many additional contributions to the equations of motion into the
volume driving force.  The amplitude is larger since these extra contributions
also have to be dissipated, but it should not exceed that of a 3D structure
function since they originally come from a 3D flow.  So in the viscous subrange
the reduction in dimensionality is not noticable and the structure functions 
coincide.

\subsection{Structure function of the vorticity field}
Another quantity of interest in 2D flows is the 
vorticity $\omega=\partial_x v-\partial_y u$ and the structure function,
\begin{eqnarray}
\hat{\Omega}(\hat{R})=
\langle|\omega(\hat{\bf x}+\hat{\bf R})-\omega(\hat{\bf x})|^2\rangle\,.
\end{eqnarray}
In 2D incompressible
turbulence squared vorticity is an additional inviscid invariant
and gives rise to an inverse cascade of energy.
In 3D a vortex stretching term  $({\boldmath\omega\cdot}{\bf \nabla}){\bf u}$ is
present that prevents a conservation of enstrophy. 
In 2D and for the normal
component of the vorticity this reduces to a normal gradient of the 
velocity field which by incompressibility is connected to the divergence
of the flow field in the surface. Thus,
for the 2D free surface flow the vorticity transport equation reads
\begin{equation}
\partial_t\omega+ (\hat{\bf u}\cdot\hat{\nabla}) \omega
= - \omega (\hat{\nabla}\cdot\hat{\bf u})+\nu\hat{\Delta}\omega
+\tilde f_{\omega}\,.
\label{vorequation}
\end{equation}
Thus, the non-vanishing divergence of the surface flow provides a kind of
additional vorticity forcing in 2D. Consequently, squared vorticity cannot be
an inviscid invariant, and no inverse cascade develops.

The vorticity structure function for the data underlying Fig.~2 is
shown in Fig.~4.  It saturates for larger separations to a
non-vanishing value.  Non-vanishing vorticity fluctuations were also observed in
the experiment \cite{Gol00} and interpreted as an indication that the observed
features are not connected with turbulent surface waves \cite{surfacewaves}.  
Note that in incompressible stationary turbulence the
second order velocity and the second order vorticity structure function are
connected by an exact relation $\Omega(R)=2\epsilon/\nu-\Delta S_2(R)$
\cite{GroMer92,SE99}.  This holds true in two and three dimensions, and has
additional terms if the flows are not incompressible.
\begin{figure}
\begin{center}
\epsfig{file=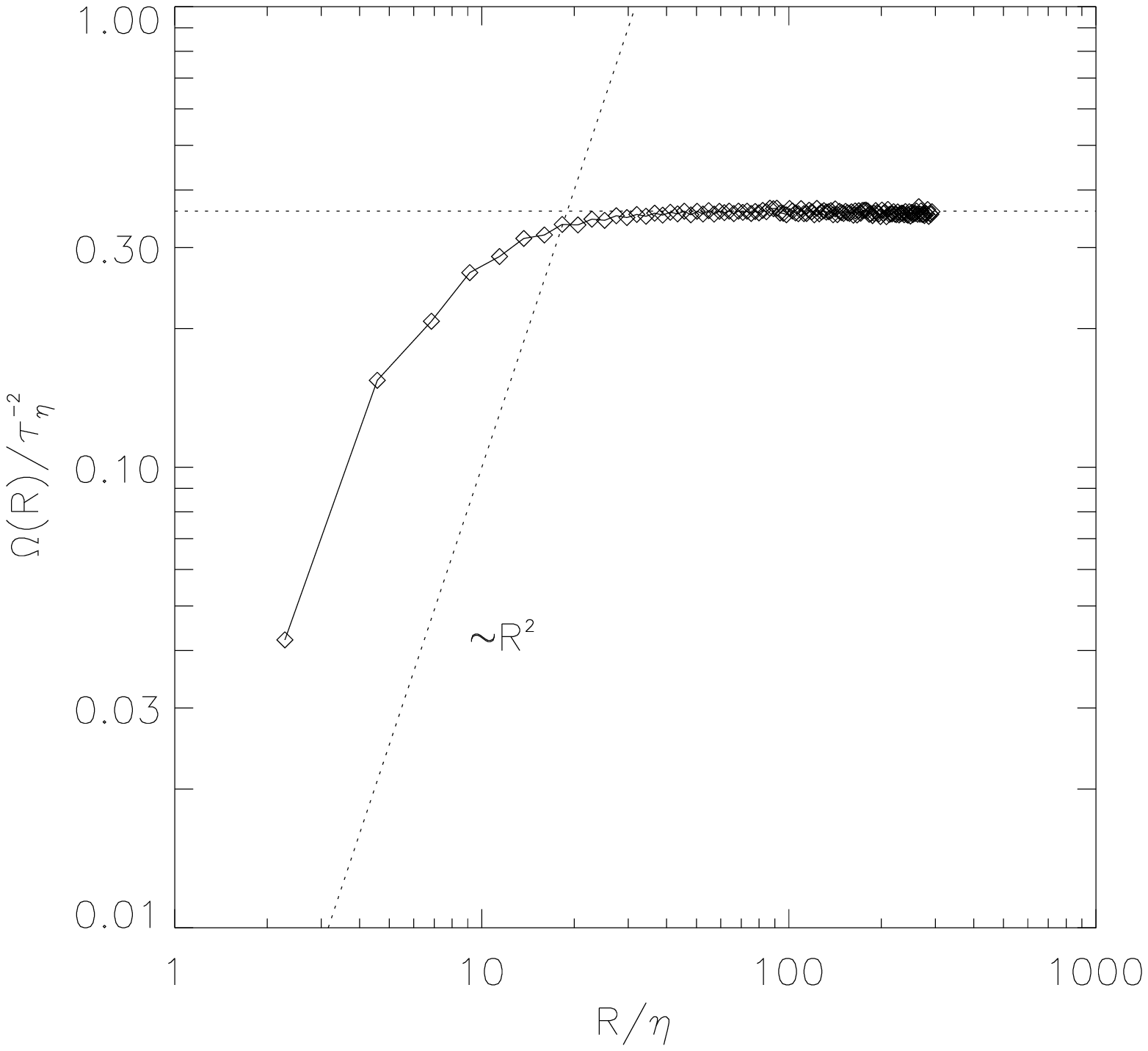,width=6.5cm}
\end{center}
\end{figure}
FIG. 4. Vorticity structure function ${\Omega}({R})/\tau_{\eta}^{-2}$ for 
the surface flow at $Re_{\lambda}=99$. The data base is the same as for 
Fig.~2.
\vspace{0.5cm}
\noindent
\begin{figure}
\begin{center}
\epsfig{file=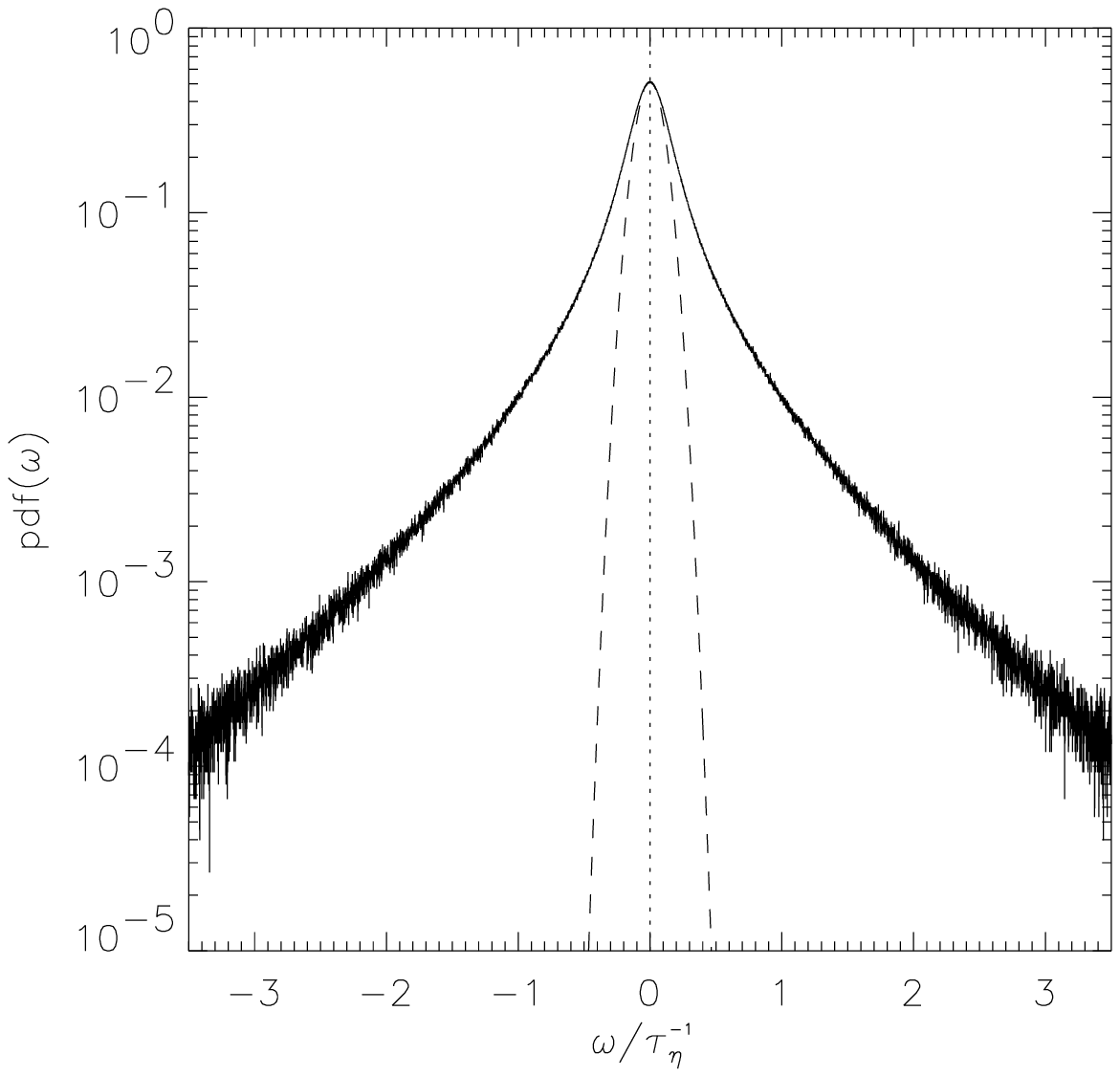,width=6.5cm}
\end{center}
\end{figure}
FIG. 5. Probability density function (pdf) of the surface vorticity component
for the flow field of  Fig.~2. For comparison a
Gaussian pdf fitted to the central part of the distribution 
is indicated as well (dashed line).
\vspace{0.5cm}
\noindent

The strong intermittency of the flow is also reflected in the probability
density function.  Fig.~5 shows that the probability density
function of the vorticity
deviates from a Gaussian distribution and has the exponentially stretched tails
that are typical for intermittent quantities.

\subsection{Divergence of the surface flow}
The property that distinguishes surface flows from incompressible
2-d flows most clearly is the divergence of the flow, which does
not vanish for the surface flow. Snapshots of the flow field,
such as in Fig.~6, clearly show the presence of sources
and sinks. 
\begin{figure}
\begin{center}
\epsfig{file=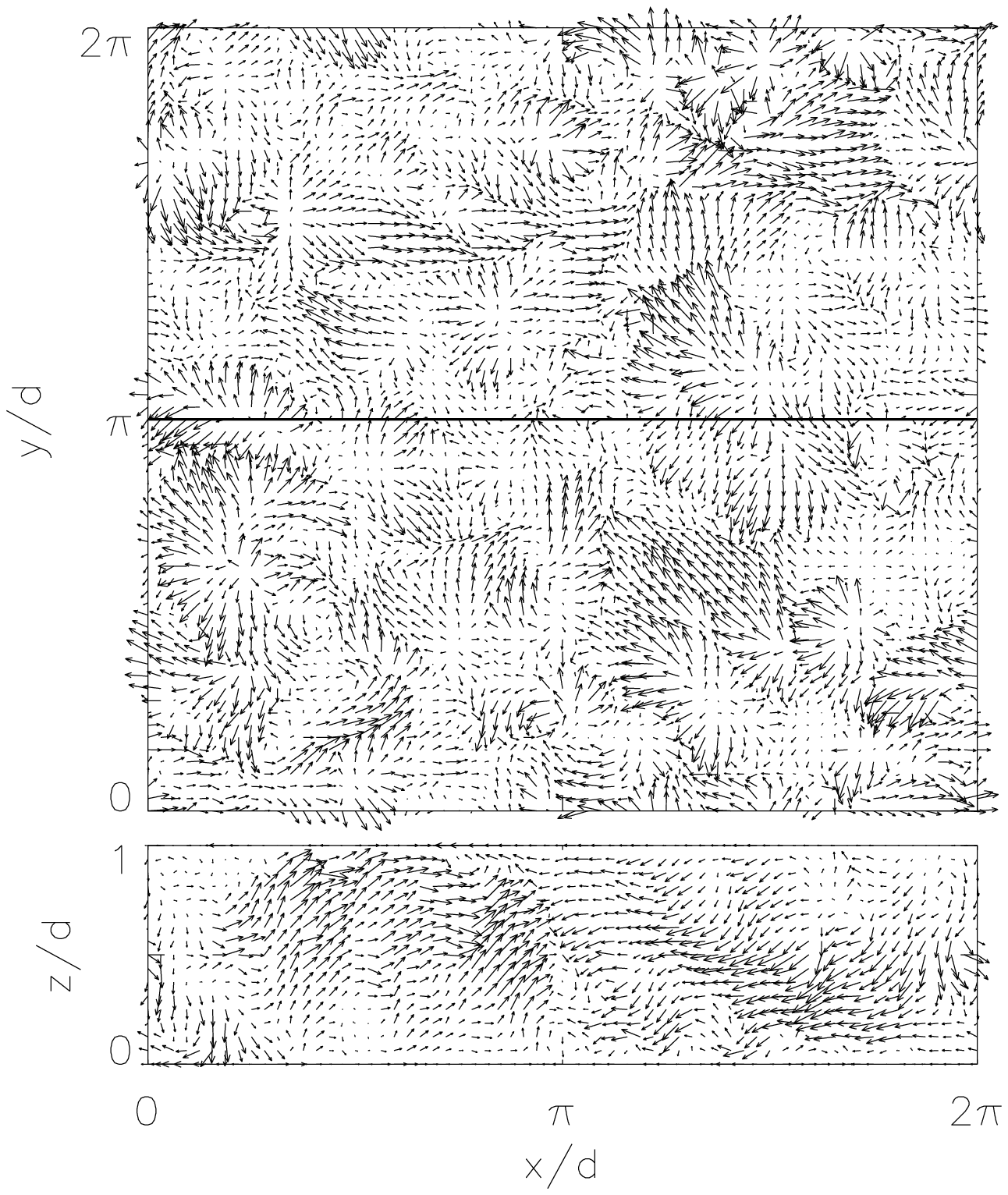,width=7cm}
\end{center}
\end{figure}
FIG. 6. A turbulent velocity field in the free-slip surface flow for $Re_{\lambda}=99$. 
The upper panel shows a vector plot of the components $u$ and
$v$ in the surface at $z/d=0$. The lower panel shows a vertical cut through the box
at the horizontal line ($y/d=\pi$) marked in the upper panel. 
Regions of rising fluid and sinking fluid in the lower panel can be connected
to sources and sinks near the solid line in the upper panel.
\vspace{0.5cm}
\noindent
A vertical slice across the flow underneath the surface
allows to connect them to up- and down-welling motions below the surface.
The corresponding contour plot of the divergence of the surface flow 
(Fig.~7)
shows randomly fluctuating patches of sources and sinks.
In the mean the flow is divergence free, 
$\langle (\hat{\nabla}\cdot\hat{\bf u})\rangle=0$,
but the root mean square value does not vanish.
Formally one can define a compressibility factor
\cite{Gaw00} 
\begin{equation}
0\le {\cal C}=\frac{\langle({\bf\nabla}\cdot{\bf u})^2\rangle}
            {\langle|{\bf\nabla}{\bf u}|^2\rangle}\le 1\,,
\label{cfactorallgemein}            
\end{equation}
which relates the mean square divergence to the mean square velocity gradient.
For the surface flow and using only the velocity components in the surface, 
this becomes
\begin{equation}
{\cal C}=\frac{\langle(\hat{\bf\nabla}\cdot\hat{\bf u})^2\rangle}
            {\langle|\hat{\bf\nabla}\hat{\bf u}|^2\rangle}\,.
\label{cfactor}            
\end{equation}
\begin{figure}
\begin{center}
\epsfig{file=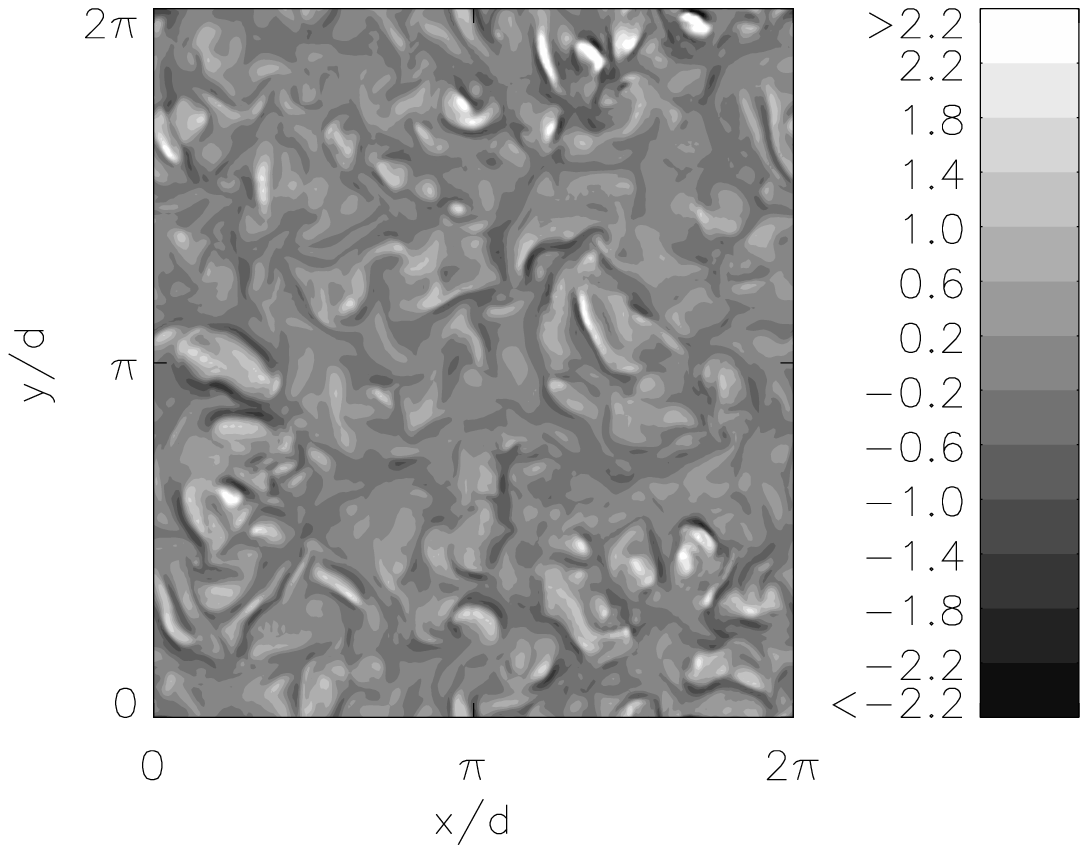,width=7cm}
\end{center}
\end{figure}
FIG. 7. Divergence $\hat{\nabla}\cdot\hat{\bf u}$ of the surface 
flow in units of the Kolmogorov time $\tau_{\eta}$. Data are the same as
for Fig.~6.
\vspace{0.5cm}
\noindent
In our simulations for $Re_{\lambda}=99$
we find ${\cal C}\approx 1/2$, in good agreement with the
Pittsburgh experiments \cite{Gol00}. The relation of the denominator in
(\ref{cfactor}) to the energy dissipation rate (\ref{epsilonsurface}) is given by
\begin{equation}
\epsilon=\nu\left[\langle|\hat{\bf\nabla}\hat{\bf u}|^2\rangle_S+
             \langle(\hat{\bf\nabla}\cdot\hat{\bf u})^2\rangle_S\right]
        =\nu\langle|\hat{\bf\nabla}\hat{\bf u}|^2\rangle_S\,(1+{\cal C})     
             \,.
\label{epsilonsurface1}            
\end{equation}
\begin{figure}
\begin{center}
\epsfig{file=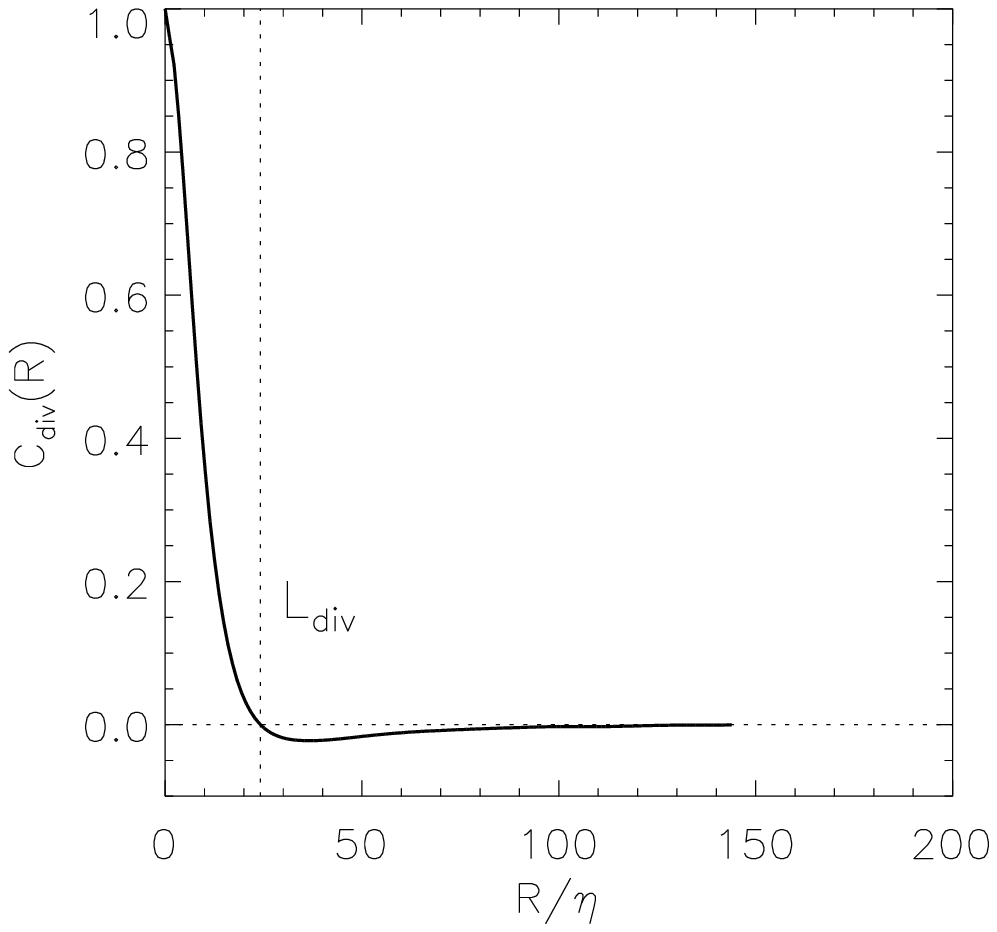,height=5cm,width=6cm}
\end{center}
\end{figure}
FIG. 8. Radially averaged correlation function of the divergence field for 
$Re_{\lambda}=99$.
The decorrelation length $L_{div}$ is indicated by the vertical dotted line.
\vspace{0.5cm}
\noindent

The mean extension of regions with similar divergence can be determined
from the correlation function,
\begin{eqnarray}
C_{div}(\hat{R})=\left\langle 
\left\lbrack \hat{\nabla}\cdot\hat{\bf u}(\hat{\bf x}) \right\rbrack\,
\left\lbrack \hat{\nabla}\cdot\hat{\bf u}(\hat{\bf x}+\hat{\bf R})
\right\rbrack 
\right\rangle\,.
\end{eqnarray}
This correlation function is shown in Fig.~8. 
The first zero of $C_{div}(\hat{R})$ defines a decorrelation length scale 
$L_{div}$;
in units of the Kolmogorov scale $L_{div}\approx 25$.
This scale fits rather well with the size of the largest patches 
in Fig.~7. 
As a consequence, the term in Eq.~(\ref{vorequation}) 
that contains the divergence of the velocity field describes 
a driving force that can be expected to be confined 
to the smaller scales in the flow.
\begin{figure}
\begin{center}
\epsfig{file=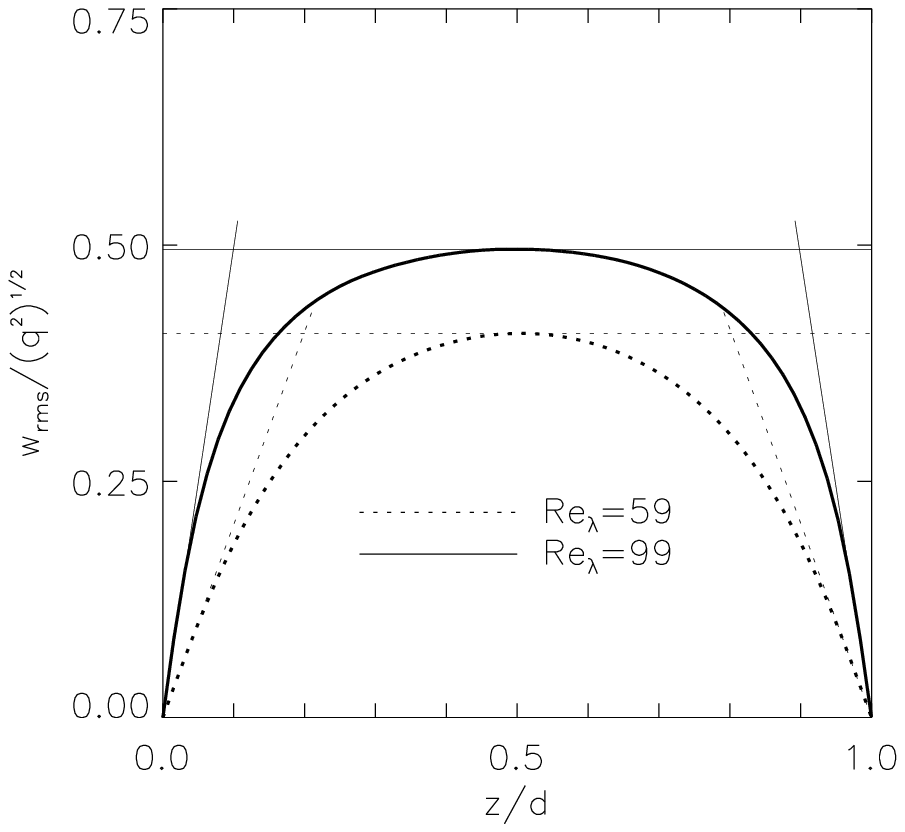,width=7cm}
\end{center}
\end{figure}
FIG. 9. Fluctuations of the normal velocity component as a function of the position 
between the surfaces. The fluctuations are normalized in units of
the square root of the mean turbulent kinetic energy
$q^2=\langle u^2\rangle_V+\langle v^2\rangle_V+\langle w^2\rangle_V$. 
A linear extrapolation from the surface up to the value in the middle
gives a boundary layer thickness 
$\delta=0.2$ for
$Re_{\lambda}=59$ and $\delta=0.1$ for $Re_{\lambda}=99$.
This is about a factor of 10 larger than the values estimated from
Eq.~(\ref{randschicht})
\vspace{0.5cm}
\noindent

\subsection{Fluctuations of the vertical velocity component}
\label{layer}
In section IIIA we already mentioned the variations of the
statistical properties with distance from the surface.
They allow to identify a surface layer in which the transition from
bulk to surface behavior takes place.
This layer is different from the 
ones near rigid walls and not connected to friction but
rather to the suppression of velocity fluctuations in wall normal
direction. Dimensional arguments allow to determine the
layer thickness $\delta$  from a balance between the 
turbulent transport of wall normal fluctuations into the boundary, 
\begin{eqnarray}
({\bf u}\cdot \nabla){\bf u} \approx w_{rms}^2 / \delta\,,
\end{eqnarray}
and the viscous dissipation of such fluctuations,
\begin{eqnarray}
\nu \Delta {\bf u} \approx \nu w_{rms}/\delta^2 \,.
\end{eqnarray}
In both cases the size of the velocity field is estimated
by the root mean square average of the wall normal velocity fluctuations, 
$w_{rms}$. Equating the two expressions gives
\begin{equation}
\delta \approx \nu/w_{rms}\,
\label{randschicht}
\end{equation}
as an estimate of the thickness. This is compared with numerical data
in Fig.~9 where the $z$ profiles of the wall--normal
fluctuations $w_{rms}$ for two values of Taylor Reynolds number
are shown. The boundary layer becomes smaller with increasing $Re_{\lambda}$,
as expected. The absolute values for the thickness of the boundary
layer can be read off from the data by linear extrapolations of the
profile slopes at the surfaces. The intersection of these straight lines
with the corresponding maximum values of $w_{rms}$ 
were used to define the boundary layer 
thickness (see Fig.~9). This gives values for $\delta$ that are
about a factor of 10 larger than the dimensional
estimate,
but consistent with its scaling behavior. 
This estimate for the thickness of a surface layer agrees with
the interval over which the scaling exponents
in (Fig.~3) change from bulk to surface values.

\section{Passive scalar transport in the free-slip surface}
\subsection{Time scales}
Experimental and numerical studies show that the particles floating
on the surface of a fluid cluster in regions with down-welling and
avoid regions with up-welling motion \cite{Ott3,Sommerer}. 
The patterns that appear have
huge density variations that are best described by fractal
scaling exponents. As an approximation to the dynamics
of particles we can study the advection of a scalar density on the
surface: it differs from true particles in that it has no inertia
(the importance of which can be reduced in experiments 
by sufficiently small and light particles, see however 
\cite{Babiano00,Balkovsky01}) 
and that it can develop larger gradients.

Allowing for the compressibility of the flow field, the 
equations for the scalar density thus are
\begin{eqnarray}
\partial_t \phi + \hat{\nabla}\cdot(\hat{\bf u} \phi)
= D \hat{\Delta} \phi + f_\phi \,,
\end{eqnarray}
where $D$ is the passive scalar diffusivity. The Prandtl number is
$Pr=\nu/D$.
The following discussion will be confined to the two-dimensional
flow, so that all gradient, divergence and Laplace operators
act on the two coordinates $x$ and $y$ only; the superscript will henceforth
be omitted. Expanding the second term in the above
expression then gives the evolution equation for the
scalar,
\begin{equation}
\partial_t \phi = - ({\bf u}\cdot \nabla) \phi
- (\nabla\cdot {\bf u}) \phi
+ D \Delta \phi + f_\phi \,.
\label{scalarequ}
\end{equation}
The input $f_\phi$ in scalar density is needed in order to 
compensate the diffusive losses. 
Since the equation for the scalar is linear, the 
natural amplitude scale for $\phi$ is set by its root mean 
square value, $\phi_{rms}=\langle\phi^2\rangle^{1/2}$. 
After dividing by $\phi_{rms}$ all terms have dimensions
of inverse time and the time scales involved can be
used to characterize the different processes.
Several of these processes also depend on the length scale $l$ over
which they are studied and so we introduce
length scale resolved characteristic times. 
All terms can be made dimensionless using the inner scales of the 
turbulent velocity field in the surface, as discussed above 
Eq.~(\ref{epsilonsurface}).
Different estimates of $\epsilon$ bring in factors of order 1
(see e.g. relation (\ref{epsilonsurface1}) and the remarks below 
Eq.~(\ref{epsilonsurface}).
 Again, we use the energy dissipation rate 
which is given by Eq.~(\ref{epsilonsurface}) to compose $\eta$, $v_{\eta}$,
and $\tau_{\eta}$.

The advection term, $({\bf u}\cdot \nabla)\phi$ in Eq.~(\ref{scalarequ})
is characterized by the advective time scale 
\begin{equation}
\tau_{adv}(l) = l/u(l) = \frac{\eta}{v_{\eta}} \left(\frac{l}{\eta}\right)^{2/3}
= \tau_{\eta} \left(\frac{l}{\eta}\right)^{2/3}
\end{equation}
where the scale resolved velocity $u(l) \approx v_{\eta} (l/\eta)^{1/3}$
in Kolmogorov theory.
The next term in Eq.~(\ref{scalarequ}) contains the divergence of the 2D surface
flow and acts like a source or sink for the scalar.  Its time scale is denoted
$\tau_{div}(l)$.  The compressibility factor relates the divergence of the flow
field to the root mean square velocity gradient [see
Eq.~(\ref{cfactorallgemein})] which is connected with the energy dissipation rate
$\epsilon$ and thus the Kolmogorov time $\tau_{\eta}$.  Numerical simulations and
experiment \cite{Gol00} indicate $\tau_{div}(l)\approx \tau_\eta/{\cal C}^{1/2}$,
where ${\cal C}$ is the compressibility factor (\ref{cfactor}) with a value of
about $1/2$.
The efficiency of diffusion clearly depends on $l$, so that the time scale
$\tau_{diff}(l)$ for diffusive smearing is $\tau_{diff}(l)\approx l^2/D$.
Finally, we have the forcing time $\tau_f=\phi_{rms}/f_\phi$, which again is
independent of spatial resolution.

In any given range of length scales, the process with the shortest time scale can
be expected to dominate.  So starting from the smallest scales we expect for an
incompressible fluid first a diffusion dominated regime, then an advection
dominated one and finally the input dominated regime.  A Batchelor regime for the
scalar is found if the diffusive regime extends beyond the Komogorov length, i.e.
$Pr\gg 1$.  In an incompressible flow, ${\cal C}=0$ and $\tau_{div}$ is infinite,
so that there is no influence from the divergence.  In the surface flows studied
here, the estimate for ${\cal C}$ indicates  that $\tau_{div}$ is very
short, of the order of the Kolmogorov time.  This implies that the advective
regime is suppressed and that the statistics of the divergence dominates.
This, finally, explains why the properties of the hydrodynamic flow do not seem
to matter too much in the analysis of the particle distribution on free surfaces
and why Ott {\it et al.}  could explain the experiments using random maps
\cite{Ott1,Ott2,Ott3,Sommerer}.

\subsection{Application of geometric scaling theory}

In order to connect the scaling of the velocity field to the scaling properties
of the scalar, we use geometric measure theory \cite{Fed69} and the
scaling ideas developed by Constantin and
Procaccia\cite{Con91,ConPro93,ProCon93}. 
 A further extension of their work allowed for a scale resolved and Prandtl
number dependent analysis \cite{GroLoh94,ES99}.  The basic idea of the
approach is to relate the fractal dimension $\delta_g^{(2)}$ of the
passive scalar concentration, i.e.  the scaling exponent (with respect to $R$)
of the Hausdorff volume $H$ of the passive scalar graph $G(B_R^{(2)})=\{({\bf
x},\phi)|{\bf x}\in B_R^{(2)},\, \phi=\phi({\bf x})\}$ taken over a
two-dimensional ball $B_R^{(2)}$ of radius $R$, to scaling properties of the
underlying turbulent flow that mixes the scalar. The following
discussion will focus on the new terms relevant to the current problem; more
details can be found in the above mentioned references 
\cite{Con91,ConPro93,ProCon93,GroLoh94} and our previous work \cite{ES99}.

The basic quantity to be calculated within geometric measure theory
is the relative Hausdorff volume of a surface of the normalized
scalar density $\tilde \phi=\phi/\phi_{rms}$, as given by 
\begin{eqnarray} 
\frac{H(G(B_R^{(2)}))}{V(B_R^{(2)})}&\sim& R^{\delta_g^{(2)}-2}\,\nonumber\\
&\le&\sqrt{1+\frac{1}{\pi} \int_{B_R^{(2)}}\,
|\nabla\tilde{\phi}|^2 
\,\mbox{d}^2{\bf x}}\,,
\label{vol}
\end{eqnarray} 
where $V(B_R^{(2)})=\pi R^2$ is the volume of a two-dimensional ball with
radius $R$. The scaling exponent of the first order scalar structure function
and fractal dimensions can be related by inequalities, which for
the analysis are assumed to be sharp
 \cite{ConPro93}. 
Using the relation $\phi \Delta \phi=\Delta\phi^2/2-|\nabla\phi|^2$
and the equation of motion (\ref{scalarequ}), the gradient under the integral
can be replaced by
\begin{equation}
|{\bf\nabla}\tilde{\phi}|^2=
-\frac{1}{2D}({\bf u}\cdot{\bf \nabla})\tilde{\phi}^2
-\frac{1}{D}\tilde{\phi}^2({\bf\nabla\cdot u})
+\frac{1}{2}\Delta\tilde{\phi}^2
+\frac{f_{\phi}\tilde{\phi}}{D\phi_{rms}}\,.
\label{gradient}
\end{equation}
With
\begin{equation}
({\bf u}\cdot{\bf \nabla})\tilde{\phi}^2
=
{\bf\nabla\cdot}\left({\bf u}\tilde{\phi}^2\right)-
\tilde{\phi}^2({\bf\nabla\cdot u})\,
\end{equation}
the first term on the right hand side can be expressed 
as a sum of two divergences.
When substituted under the integral 
in Eq.~(\ref{vol}) the Hausdorff volume becomes
\end{multicols}
\begin{equation}
\frac{H(G(B_R^{(2)}))}{V(B_R^{(2)})}
\le\sqrt{1+\frac{1}{\pi} \int_{B_R^{(2)}}\, 
\left\{
\frac{1}{2D}[-{\bf\nabla\cdot}({\bf u}\tilde{\phi}^2)-
\tilde{\phi}^2({\bf\nabla\cdot u})
+D\Delta\tilde{\phi}^2]
+\frac{f_{\phi}\tilde{\phi}}{D\phi_{rms}}
\right\} 
\,\mbox{d}^2{\bf x}}\,.
\label{vol1}
\end{equation} 
\begin{multicols}{2}
The four integrals are denoted $I_1$ through $I_4$ and analyzed
separately.
The analysis of the three
integrals $I_1$, $I_3$, and $I_4$ proceeds as in the previous 
applications to two-dimensional scalar advection \cite{ES99}. 
In particular, application of Gauss' theorem and the 
Cauchy-Schwartz inequality connects the first integral
to the longitudinal structure function of the surface
velocity field $S_{2}^L(R)$,
\begin{equation}
I_1 \le \frac{\sqrt{F_{\phi}}}{D} R \sqrt{S_{2}^L(R)} \,.
\label{i1eq}
\end{equation}
$F_{\phi}$ is the flatness of the
passive scalar,
\begin{eqnarray}
F_\phi =\langle\phi^4\rangle /\langle\phi^2\rangle^2=
\langle\tilde{\phi}^4\rangle\,. 
\end{eqnarray}
If the correlations of $\phi$ decay rapidly this is
essentially the volume average $\langle \phi^4 \rangle_V
/\langle \phi^2\rangle_V^2$. For a Gaussian velocity
field, $F_\phi=3$. 
Experiments and numerical simulations indicate strong ramp and cliff
structures in the scalar field and thus some deviation from the Gaussian
distribution \cite{Britz,Holzer,Mydlarski,Celani},  implying
a scale dependence of $F_\phi$. However, we here restrict ourselves
to first and second order correlations where intermittency corrections 
to the
classical Kolmogorov--Obukhov--Corrsin scaling are expected to be small
and work with a constant $F_\phi$.

Exploiting the statistical stationarity of the passive scalar dynamics
the fourth term, which contains the driving of the passive scalar,
can be expressed as
\begin{eqnarray} 
\label{i4eq} 
I_4&=&\frac{1}{\pi}\int_{B_R^{(2)}}\,\frac{f_{\phi}\tilde{\phi}} 
{D \phi_{rms}}\,\mbox{d}^2{\bf x}\,,\nonumber\\ 
&=&\frac{R^2}{D \phi_{rms}^2}\,\frac{1}{\pi R^2}\int_{B_R^{(2)}}
f_{\phi}\phi\,\mbox{d}^2{\bf x}\,,\nonumber\\             
&=&Pr \frac{\tau_{\eta}}{\tau_f}\tilde{R}^2\,,  
\end{eqnarray} 
where $\tilde{R}=R/\eta$ is the radius of the disk in units of the 
Kolmogorov length. Using Gauss' theorem and Green's formula 
it can be shown \cite{ES99} that 
the third term is subdominant
compared to the fourth,
\begin{eqnarray}
I_3\le 2\sqrt{I_4} \propto\tilde{R}\,,
\end{eqnarray}
and hence can be omitted in the following.

Finally, we come to the new term, $I_2$, which
contains the divergence of the velocity field.
Application of the Cauchy-Schwartz inequality gives
\begin{eqnarray}
I_2&=&-\frac{1}{2\pi D} \int_{B_R^{(2)}}\tilde{\phi}^2({\bf\nabla\cdot u})
\,\mbox{d}^2{\bf x}\,,\nonumber\\
&\le&
\frac{R^2}{2 D}\sqrt{\int_{B_R^{(2)}}
\frac{\tilde{\phi}^4}{V(B_R^{(2)})}\,\mbox{d}^2{\bf x}}
\;\sqrt{\int_{B_R^{(2)}}\frac{({\bf\nabla\cdot u})^2}{V(B_R^{(2)})}
\,\mbox{d}^2{\bf x}}\,,\nonumber\\
&=&\frac{\sqrt{F_{\phi}} R^2}{2 D} 
\,\langle({\bf\nabla\cdot u})^2\rangle^{1/2}\,,\nonumber\\
&=&\frac{\sqrt{F_{\phi}} Pr \tilde{R}^2}{2} 
\,\langle(\widetilde{\bf\nabla\cdot u})^2\rangle^{1/2}\,,
\label{I2div}
\end{eqnarray}
where the root mean square of the divergence is measured in units of the
Kolmogorov time $\tau_{\eta}$. 
The derivatives in the divergence term can be estimated
from above using
$\langle({\bf\nabla\cdot u})^2\rangle\le\langle|{\bf\nabla u}|^2\rangle$
[see (\ref{cfactor})].
In the following calculation this bound is not needed and the divergence
fluctuations in  
Eq.~(\ref{I2div}) can be taken directly from the numerical simulations.

Combining (\ref{vol1}), (\ref{i1eq}), and (\ref{i4eq}) 
we arrive at 
an inequality for
the fractal dimension $\delta^{(2)}_g$ of the passive scalar graph 
and thus via $\delta^{(1)}_g=\delta^{(2)}_g-1$ \cite{ConPro93} at a fractal
dimension for the
constant level sets $\phi_0=\phi({\bf x})$,
\begin{eqnarray} 
\label{fracdim} 
\delta^{(1)}_g-1\le\frac{\mbox{d}}{\mbox{d}\,\ln\tilde{R}}
\ln h(\tilde{R})\,, 
\end{eqnarray}
with
\end{multicols}
\begin{equation}
h(\tilde{R}) = 
\sqrt{1+
\sqrt{F_{\phi}} Pr\, \tilde{R}\sqrt{\tilde{S}_{2}^L}
+\frac{\sqrt{F_{\phi}} Pr \tilde{R}^2}{2} 
\,\langle(\widetilde{\bf\nabla\cdot u})^2\rangle^{1/2}
+Pr \frac{\tau_{\eta}}{\tau_f} \tilde{R}^2
}\;.
\label{hvonr}
\end{equation} 
\begin{multicols}{2} 
\noindent
$\tilde{S}_{2}^L=S_{2}^L/v_{\eta}^2$ is the longitudinal second order
structure function in units of the Kolmogorov velocity.  Obviously, if the
last two terms under the integral dominate, then $h(\tilde{R})\approx \tilde{R}$
and $\delta^{(1)}_g=2$, implying a surface filling distribution of the scalar.
For sufficiently large scales $\tilde{R}$ or large Prandtl numbers this is
always the case. On the other
hand, if the second term with the velocity structure function dominates, say
$\tilde{S}^L_{2} \approx \tilde{R}^\gamma$, then $\delta^{(1)}_g =
3/2+\gamma/4$.  Thus for the usual Kolmogorov scaling,
$\tilde{S}^L_{2}\sim\tilde{R}^{2/3}$ and $\delta^{(1)}_g = 5/3$.

The Prandtl-number dependence of the fractal dimension can be studied using 
as input the velocity correlations functions from our numerical simulations.
Besides $Pr$ also the prefactor $\tau_{\eta}/\tau_f$, a measure of the strength
of the scalar driving, is a free parameter in (\ref{hvonr}).  The results for
two different values of $\tau_{\eta}/\tau_f$ are shown in Fig.~10.  For
many values of $Pr$ a fractal dimension $\delta^{(1)}_g<2$ is observed.  If the
term $\tau_{\eta}/\tau_f$ becomes large, either because of a small $\tau_f$
(strong driving) or large $\tau_{\eta}$ (weak transport to smaller scales), the
fractal dimension approaches that of a space filling fractal, $\delta^{(1)}
\approx 2$.

In the experiments of Sommerer \cite{Sommerer} a fractal dimension
$\delta^{(1)}_g$ between 1.28 and 1.43 (denoted $D_2$) was found.  We find these
values only in the transitional region, before the inertial range is developed.
The observations are consistent with Eq.~(\ref{hvonr}) since it only provides an
upper bound and the observed values are smaller, indeed.  Further comparisons
between experiment and theory, using e.g.  measured velocity correlation
functions in $(\ref{hvonr})$, would be more than welcome.  Some experiments are in
preparation \cite{Gol00}.

\begin{figure}
\begin{center}
\epsfig{file=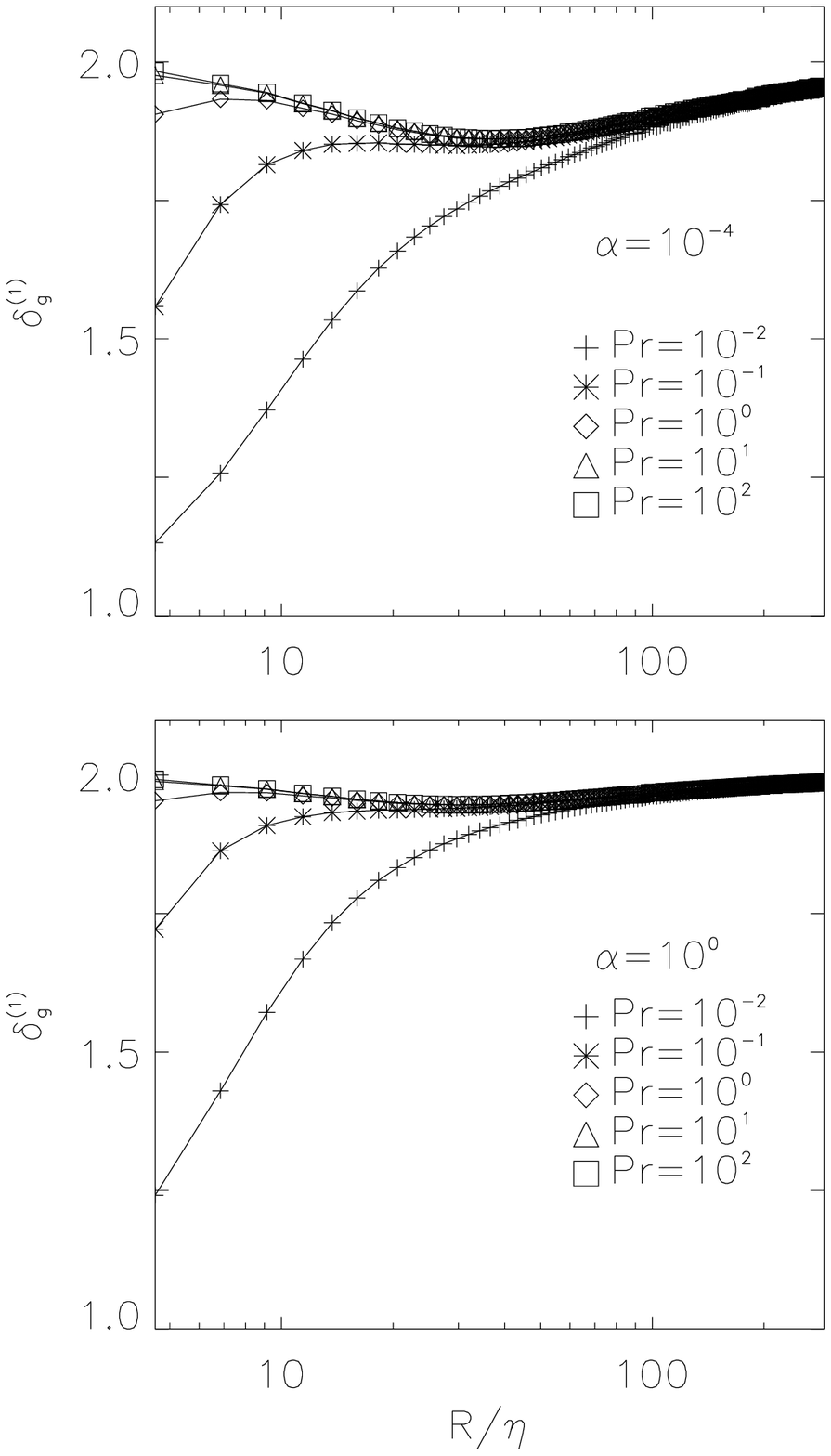,width=7cm}
\end{center}
\end{figure}
FIG. 10. Fractal dimension $\delta_g^{(1)}$ for passively advected scalars for different
values of the parameter $\alpha=\tau_{\eta}/\tau_f$ and the 
Prandtl number $Pr$. The 
underlying turbulent velocity field is the surface flow for $Re_{\lambda}=99$
as shown in Fig.~2 and $F_{\phi}=3$.
\vspace{0.5cm}
\noindent

\section{Summary}

The surface flows studied here are intermediate between two- and
three-dimensional flows.  They are confined to a surface, but their statistical
properties are strongly influenced, even dominated by the 3D volume turbulence.
The flow field in the surface can exchange energy and vorticity with the bulk,
so that neither energy nor enstrophy are conserved quantities in the Eulerian,
undriven limit.  Moreover, in addition to large scale
forces that maintain the 3D flow the surface flow is driven by small scale 
perturbations that come from
transversal pressure variations and local gradients in normal velocity.  As a
result the scaling properties of the flow are essentially that of 3D turbulence,
with an energy cascade in the inertial regime.  The scaling exponents of the
velocity structure function are slightly larger than those of bulk 3D turbulent
flows, indicating larger intermittency effects.  We also observed a 2/3
difference in the amplitude of the structure functions between surface and bulk
in the inertial regime.  Several of the observed characteristics of the surface
flow are in agreement with the measurements of the Pittsburgh group
\cite{Gol00}.

We have also discussed the scaling properties of a scalar advected
by the surface flow and have identified different scaling regimes.
It seems that very often the dynamics induced by the 
divergence of the flow field is the fastest process, and that
the advective properties of the flow are subdominant. This might
explain why random mappings could succesfully be applied to the
modelling of the particle distributions 
\cite{Ott1,Ott2,Ott3,Sommerer}, but a more detailed comparison between
theory and experiment is clearly needed.

An important characteristic quantity of the surface flows is the
compressibility factor ${\cal C}$.
The numerical simulations and the Pittsburgh experiment using a 
vertically oscillating grid indicate both ${\cal C}\approx 1/2$.
With a stable stratification of the fluid below the surface 
that reduces vertical fluctuations it might be possible to 
achieve smaller values of ${\cal C}$. This should open up the 
possibility to study the effects of compressibility over
a larger range of ${\cal C}$, both in connection with the
intermittency contributions to the scaling exponents
and with the scalar dynamics in surface flows.

\subsection*{Acknowledgments}
BE would like to thank Walter Goldburg for raising the problem of surface flows and
Luca Biferale for a discussion of time scales. We thank
Rainer Friedrich and Detlef Lohse for pointing us to supersonic
compressible turbulence. Both of us would like to thank
the Institute for Theoretical Physics at Santa Barbara for hospitality.  This
work was supported in part by National Science Foundation under Grant No.
PHY94-07194, and the EU within the CARTUM project and the 
`Nonideal turbulence'-Research Training Network HPRN-CT-2000-00162.  
The numerical simulations
were done on a Cray T-90 at the John von Neumann--Institut f\"ur Computing at the
Forschungszentrum J\"ulich and we are grateful for their support.

\end{multicols}

\references
\bibitem{Stommel} H. Stommel, J. Marine Res. {\bf 8}, 199 (1949).

\bibitem{Davis} R. E. Davis, Annu. Rev. Fluid Mech. {\bf 23}, 43 (1991).

\bibitem{Ott1} L. Yu, E. Ott, and Q. Chen,
               Phys. Rev. Lett. {\bf 65}, 2935 (1990).

\bibitem{Ott2} T. Antonsen, A. Namenson, E. Ott, and J.C. Sommerer,
               Phys. Rev. Lett. {\bf 75}, 3438 (1995).

\bibitem{Ott3} A. Namenson, T. Antonsen, and E. Ott,
               Phys. Fluids {\bf 8}, 2426 (1996).

\bibitem{Sommerer} J.C. Sommerer, Phys. Fluids {\bf 8}, 2441 (1996).

\bibitem{Saich97} V. I. Klyatskin and A. I. Saichev, JETP {\bf 84}, 716 (1997).

\bibitem{Saich98} A. I. Saichev and I. S. Zhukova, Lecture Notes in Physics
                  {\bf 511}, 353 (Springer, Berlin 1998).

\bibitem{ES99} B. Eckhardt and J. Schumacher, Phys. Rev. E {\bf 60},
               4185 (1999).

\bibitem{Mar98} B. K. Martin, X. L. Wu, W. I. Goldburg, and M. A. Rutgers,
                Phys. Rev. Lett. {\bf 80}, 3964, (1998).

\bibitem{Riv98} M. Rivera, P. Vorobieff, and R. E. Ecke,
                Phys. Rev. Lett. {\bf 81}, 1417 (1998).

\bibitem{Rut98} M. A. Rutgers, Phys. Rev. Lett. {\bf 81}, 2244 (1998).

\bibitem{Gol00} W. I. Goldburg, J. R. Cressman, Z. V\"or\"os, B. Eckhardt,
                and J. Schumacher, {\em Turbulence in a free surface},
                eprint CD.nlin/0008020 (2000).

\bibitem{Frisch} U. Frisch, {\em Turbulence, The legacy of A. N. Kolmogorov},
                 (Cambridge University Press, Cambridge 1995).

\bibitem{See96}  N. Seehafer, E. Zienicke, and F. Feudel,
                 Phys. Rev. E {\bf 54},  2863  (1996).

\bibitem{Schu00} J. Schumacher and B. Eckhardt, Europhys. Lett. {\bf 52}, 627 
                 (2000).

\bibitem{benzi} R. Benzi, S. Ciliberto, R. Trippicone,
C. Baudet, F. Massaioli, and S. Succi, Phys. Rev. E {\bf 48} , R29 (1993)

\bibitem{Gaw00} K. Gaw\c{e}dzki and M. Vergassola, Physica D {\bf 138}, 63 (2000).   

\bibitem{Celani99} A. Celani, A. Lanotte, and A. Mazzino,  
                   Phys. Rev. E {\bf 60}, R1138 (1999).

\bibitem{Porter98} D. H. Porter, P. R. Woodward, and A. Pouquet,
                   Phys. Fluids {\bf 10}, 237 (1998). 

\bibitem{Pety00} J. Pety and \'{E}. Falgarone, Astron. Astrophys. {\bf 356},
                 279 (2000).                                    

\bibitem{surfacewaves} V.E. Zakharov, V.S. L'vov, and G. Falkovich, 
{\em Kolomogorov spectra of turbulence I: Wave turbulence}
(Springer-Verlag, Berlin 1992).

\bibitem{GroMer92} S. Grossmann and P. Mertens, Z. Phys. B {\bf 88}, 
                   105 (1992). 

\bibitem{SE99} J. Schumacher and B. Eckhardt , Phys. Plasmas {\bf 6}, 
               3477 (1999).

\bibitem{Babiano00} A. Babiano, J. H. E. Cartwright, O.Piro, and A. 
                    Provenzale, Phys. Rev. Lett. {\bf 84}, 5764 (2000).

\bibitem{Balkovsky01} E. Balkovsky, G. Falkovich, and A. Fouxon,
                    Phys. Rev. Lett. {\bf 86}, 2790 (2001)
.

\bibitem{Fed69} H. Federer, {\it Geometric Measure Theory}, (Springer, 
                Berlin) 1969. 

\bibitem{Con91} P. Constantin, I. Procaccia, and K. R. Sreenivasan,  
                Phys. Rev. Lett. {\bf 67}, 1739 (1991). 

\bibitem{ConPro93} P. Constantin and I. Procaccia, Phys. Rev. E {\bf 47}, 3307 
                  (1993). 

\bibitem{ProCon93} I. Procaccia and P. Constantin, Europhys. Lett. {\bf 22},
                   689 (1993).
                   
\bibitem{GroLoh94} S. Grossmann and D. Lohse, Europhys. Lett. {\bf 27}, 
                   347 (1994). 

\bibitem{Britz}   K. R. Sreenivasan, R. A. Antonia, and D. Britz,
                  J. Fluid Mech. {\bf 94}, 745 (1979).
                                                       
\bibitem{Holzer}  M. Holzer and E. D. Siggia, Phys. Fluids {\bf 6}, 1820 (1994).
                  
\bibitem{Mydlarski} L. Mydlarski and Z. Warhaft, J. Fluid Mech. {\bf 358},
                    135 (1998).
                    
\bibitem{Celani} A. Celani, A. Lanotte, A. Mazzino, and M. Vergassola, 
                 Phys. Rev. Lett. {\bf 84}, 2385 (2000).
\end{document}